\documentclass[12pt]{book}

\usepackage[dvips]{graphicx,color}
\usepackage{makeidx,physics,cosmology}

\makeauthorindex
\makeindex

\BookTitle{Frontier in Astroparticle Physics and Cosmology}
\CopyRight{\copyright 2004 by Universal Academy Press, Inc.}

\begin{document}

\BookTitle{\itshape Frontier in Astroparticle Physics and Cosmology}
\CopyRight{\copyright 2004 by Universal Academy Press, Inc.}
\pagenumbering{arabic}

\chapter{
Conduction and Turbulent Mixing in Galaxy Clusters}

\author{%
Ramesh NARAYAN \\
{\it Harvard-Smithsonian Center for Astrophysics, 60 Garden Street,
Cambridge, MA 02138, USA}\\
Woong-Tae KIM \\
{\it Astronomy Program, SEES, Seoul National University, Seoul 151-742, 
Korea}\\
{\it Harvard-Smithsonian Center for Astrophysics, 60 Garden Street,
Cambridge, MA 02138, USA}
}
%
%
\AuthorContents{R.\ Narayan and W.-T.\ Kim} 
\AuthorIndex{Narayan}{R.} 
\AuthorIndex{Kim}{W.-T.}

\section*{Abstract}

We discuss hydrostatic models of galaxy clusters in which heat
diffusion balances radiative cooling.  We consider two different
sources of diffusion, thermal conduction and turbulent mixing,
parameterized by dimensionless coefficients, $f$ and $\alpha_{\rm
mix}$, respectively.  Models with thermal conduction give reasonably
good fits to the density and temperature profiles of several cooling
flow clusters, but some clusters require unphysically large values of
$f>1$.  Models with turbulent mixing give good fits to all clusters,
with reasonable values of $\alpha_{\rm mix} \sim 0.01-0.03$.  Both
types of models are found to be essentially stable to thermal
perturbations.  The mixing model reproduces the observed scalings of
various cluster properties with temperature, and also explains the
entropy floor seen in galaxy groups.

\section{Introduction}

For many years, it was thought that the strong X-ray emission observed
in the cores of rich galaxy clusters results in a cooling flow in
which gas settles in the gravitational potential and drops out as cold
condensations \cite{NK_fab94}. Mass inflow rates were estimated to be
$\sim10^2-10^3M_\odot$ yr$^{-1}$ in some clusters.  However, recent
X-ray observations with {\it Chandra} and {\it XMM-Newton} have found
very little emission from gas cooler than about one-third of the
virial temperature \cite{NK_pet01,NK_pet03}, suggesting that some
heating source must prevent gas from cooling below this
temperature. Candidate heating mechanisms include (1) energy injection
from a central active galactic nucleus (AGN)
\cite{NK_cio01,NK_chu02,NK_bru02,NK_kai03}, and (2) diffusive
transport of heat from the outer regions of the cluster to the center
via conduction \cite{NK_tuc83,NK_bre88,NK_nar01,NK_voi02,NK_zak03} or
turbulent mixing \cite{NK_cho03,NK_kim03b,NK_voi04}.

Heating by a central AGN is an attractive idea since many cooling flow
clusters show radio jets and lobes that are apparently interacting
with the cluster gas \cite{NK_beg01}.  The power associated with the
jets is often comparable to the total X-ray luminosity of the cluster.
However, there are some difficulties with this model.  Observations
reveal that radio lobes are surrounded by X-ray-bright shells of
relatively cool gas \cite{NK_sch02}, which is a little surprising if
this gas is being heated by the bubble.  In addition, if the heating
rate (per unit volume) of the gas by the AGN varies as $\rho^\alpha$,
thermal stability requires $\alpha>1.5$ \cite{NK_zak03}; such a
heating law does not seem natural.  (Stability is not an issue if AGN
heating is episodic \cite{NK_kai03}).  Finally, no good correlation is
seen between the AGN radio luminosity and the X-ray cooling rate
\cite{NK_voi04}.

Since the cooling cores of clusters have a lower temperature than the
rest of the cluster, diffusive processes can bring heat to the center
from the outside, provided the diffusion coefficient is large enough.
An ordered magnetic field would strongly suppress cross-field
diffusion of thermal electrons, and this argument has been
traditionally invoked for ignoring thermal conduction.  However, if
the field lines are chaotically tangled over a wide range of length
scales, the isotropic conduction coefficient $\kappa_{\rm cond}$ can
be as much as a few tens of per cent of the Spitzer value $\kappa_{\rm
Sp}$ \cite{NK_nar01,NK_cha03}, which may be sufficient to supply the
necessary heat to the cluster core.  Turbulent mixing is another
diffusive process that can transport energy efficiently to the center
\cite{NK_cho03}.  The turbulence might be sustained by the infall of
small groups or subclusters, the motions of galaxies [K. Makishima,
this conference], or energy input from AGNs \cite{NK_dei96,NK_ric01}.
The diffusion coefficient required to balance radiative cooling is
typically $\kappa_{\rm mix}\sim 1-6\;\rm kpc^2\;Myr^{-1}$, which is
similar to values inferred from observations of turbulence in clusters
\cite{NK_kim03b,NK_voi04}.

In a series of papers \cite{NK_zak03,NK_kim03b,NK_kim03a}, we have
studied equilibrium models of galaxy clusters with thermal conduction
and turbulent mixing.  We summarize here the main results of this
work.

\section{Model}\label{NK_model}

We assume that the hot gas in a galaxy cluster is in hydrostatic
equilibrium and that it maintains energy balance between radiative
cooling and diffusive heating,
\begin{equation}
\frac{1}{\rho}\nabla P = -\nabla \Phi,
\qquad  
\nabla\cdot\mathbf{F} = -j,
\end{equation}
where $P$ is the thermal pressure, $\rho$ is the density, $\Phi$ is
the gravitational potential, $\mathbf{F}$ is the local diffusive heat
flux, and $j$ is the radiative energy loss rate per unit volume.  For
$kT{\lower.5ex\hbox{$\; \buildrel > \over \sim \;$}}2$keV, $j$ is
dominated by free-free emission, while for lower temperatures it is
mostly due to line cooling.

We consider two diffusive processes: thermal conduction and turbulent
mixing.  In the case of the former, the heat flux is proportional to
the temperature gradient.  In the case of the latter, turbulent
motions cause gas elements with different specific entropies to move
around and mix with one another, causing a heat flux proportional to
the entropy gradient.  Thus, we write the net heat flux as
\begin{equation} \label{NK_Flux}
\mathbf{F}=-\kappa_{\rm cond}\nabla T -\kappa_{\rm mix}\rho T\nabla s,
\;\;\; \kappa_{\rm cond}=f\kappa_{\rm Sp},
\;\;\; \kappa_{\rm mix}=\alpha_{\rm mix}c_sH_p,
\end{equation}
where $T$ is the temperature, and $s$ is the specific entropy.  We
assume that the conductivity $\kappa_{\rm cond}$ is a fraction $f$ of
the Spitzer value $\kappa_{\rm Sp}$ in an unmagnetized plasma, and the
mixing coefficient $\kappa_{\rm mix}$ is a fraction $\alpha_{\rm mix}$
of the product of the sound speed $c_s$ and pressure scale height
$H_p$. We take $H_p\approx(r_c^2+r^2)^{1/2}$, where $r$ is the local
radius and $r_c$ is the core radius \cite{NK_zak03}, and set
$s=c_v\ln(P\rho^{-\gamma})$, where $c_v$ is the specific heat at
constant volume and $\gamma=5/3$ is the adiabatic index.  For
simplicity, we have considered models with either pure conduction or
pure mixing.

\section{Results}\label{NK_res}

\begin{figure}[p]
  \begin{center}
  \includegraphics[height=18pc,width=16pc]{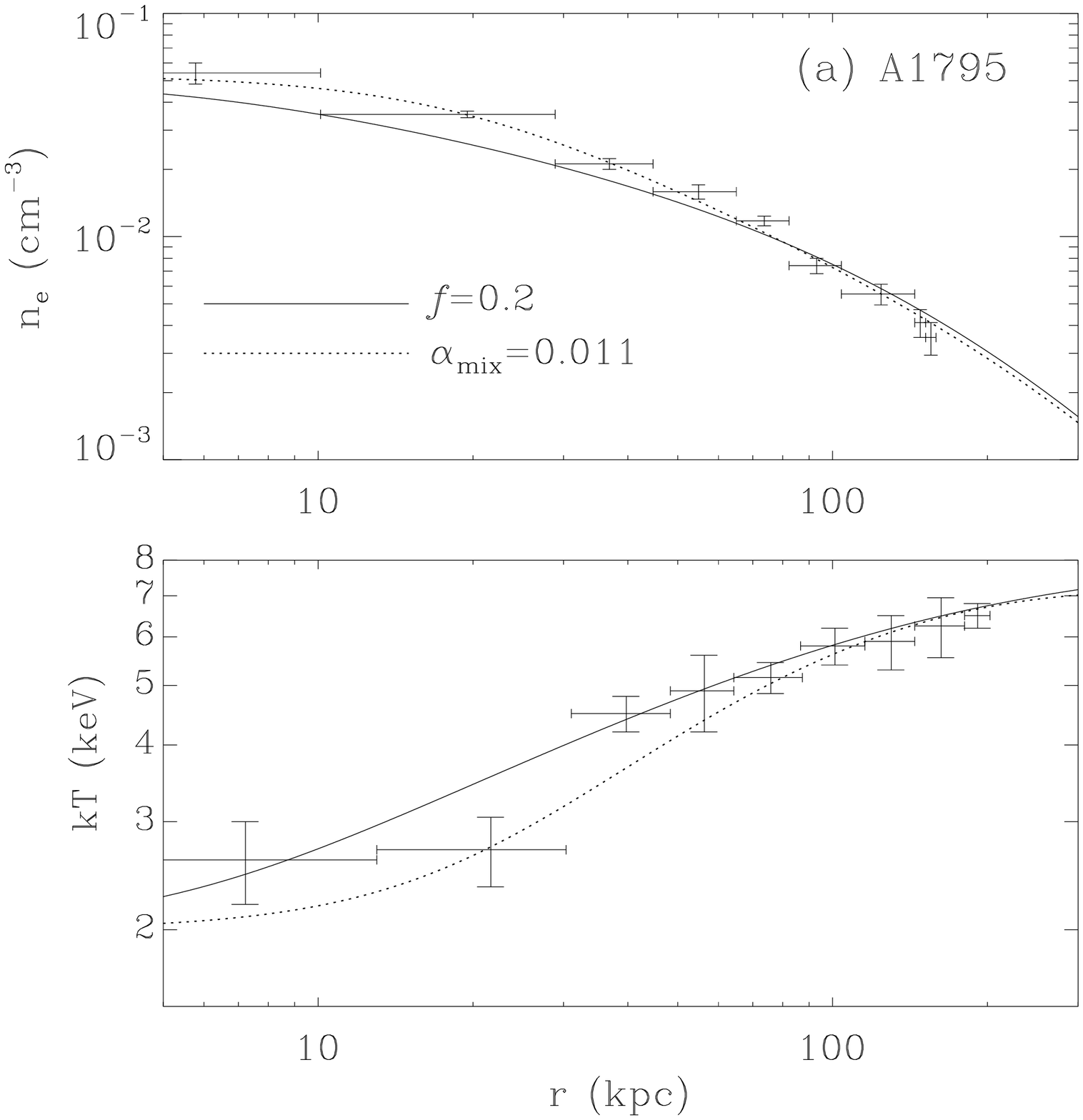}
  \includegraphics[height=18pc,width=16pc]{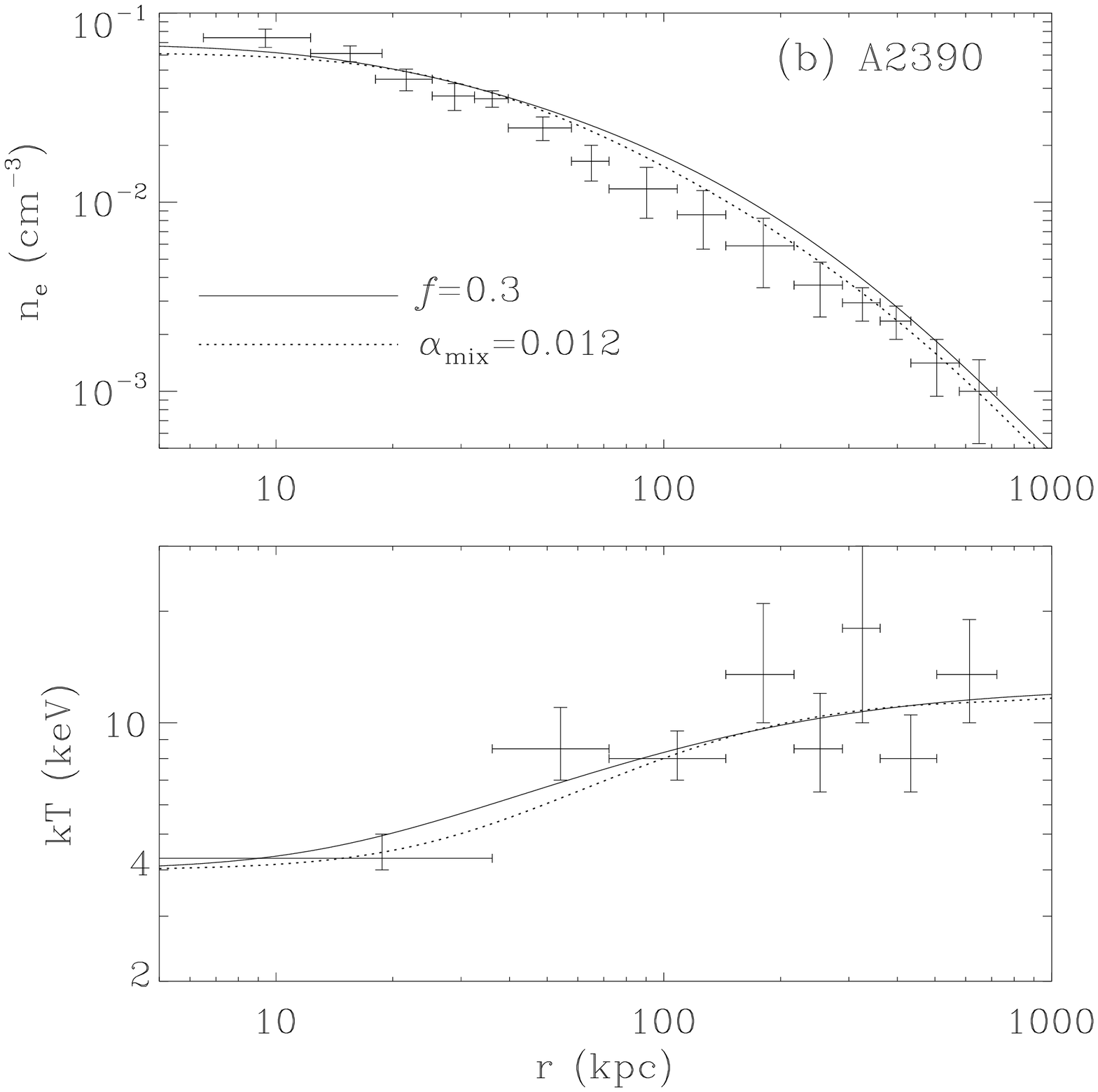}

  \vspace{1.cm} \includegraphics[height=18pc,width=16pc]{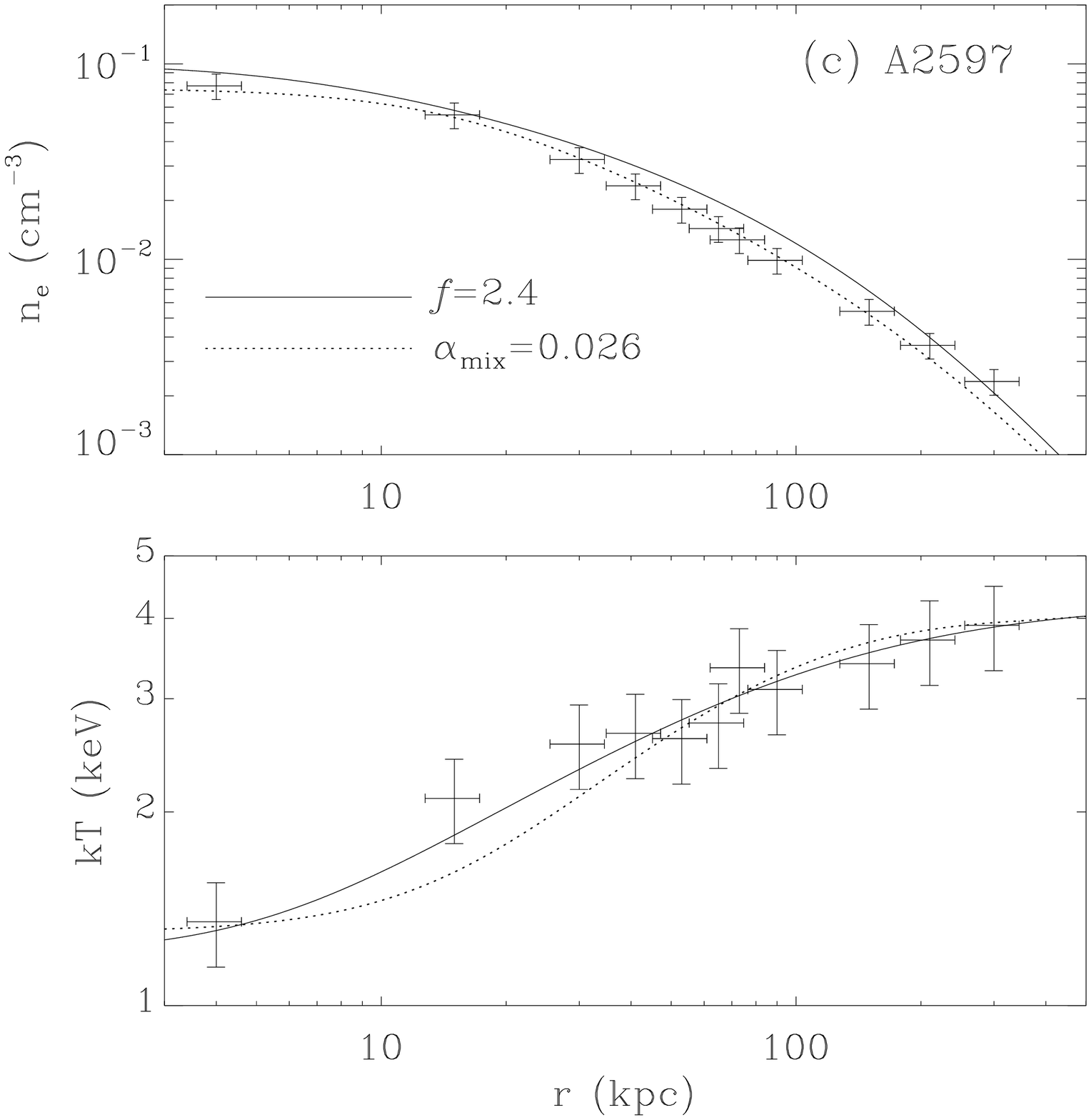}
  \includegraphics[height=18pc,width=16pc]{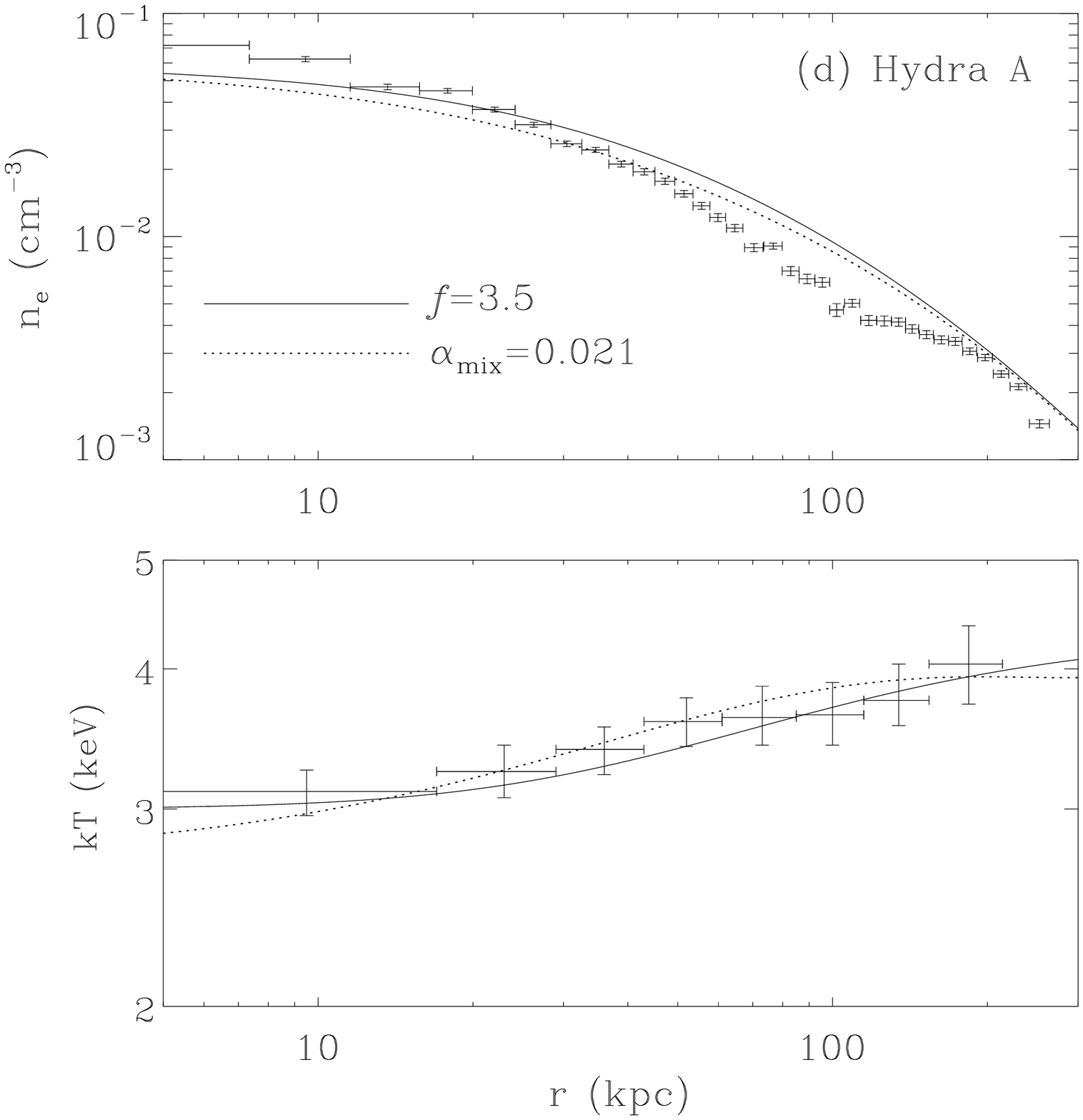} \end{center}
  \caption{Observed and modeled profiles of electron number density
  and temperature for ({\it a}) A1795, ({\it b}) A2390, ({\it c})
  A2597, and ({\it d}) Hydra A.  The data are from {\it Chandra}.  The
  solid and dotted lines represent best-fit models based on pure
  thermal conduction and pure turbulent mixing, respectively.
  $H_0=70\,{\rm km\,s^{-1}\,Mpc^{-1}}$, $\Omega_M=0.3$, and
  $\Omega_\Lambda=0.7$ have been adopted.  While the conduction model
  requires unphysically large values of $f>1$ for A2597 and Hydra A,
  the mixing model gives good fits to all four clusters with
  reasonable values of $\alpha_{\rm mix}\sim0.01-0.03$.  }
\end{figure}

We integrate the basic equations described above to calculate the
radial profiles of the electron number density $n_e(r)$ and
temperature $T(r)$.  For each cluster, we assume that the observed gas
temperature $T_{\rm obs}$ in the region outside the cooling core is
the virial temperature and use this to determine the gravitational
potential, assuming an NFW distribution for the dark matter
\cite{NK_mao97,NK_afs02}.  We also use $T_{\rm obs}$ as a boundary
condition for the gas at large radius.  We vary the central density
$n_e(0)$ and temperature $T(0)$, along with either $f$ (for the
conduction model) or $\alpha_{\rm mix}$ (for the mixing model), to
find the solution that best fits the observed density and temperature
distributions of the cluster.

We have analyzed ten clusters (A1795, A1835, A2052, A2199, A2390,
A2597, Hydra A, RX J1347.5$-$1145, Sersic 159-03, and 3C 295) for
which high resolution data are available.  Figure 1 shows the results
of the model fitting for four of these clusters. Solid lines indicate
the best-fit conduction models, while dotted lines show the best-fit
mixing models.  Overall, both models explain the observed data
reasonably well.

Of the ten clusters, five (A1795, A1835, A2199, A2390, RX
J1347.5$-$1145) are well described by a pure conduction model with
$f\sim0.2-0.4$, while the other five (A2052, A2597, Hydra A, Sersic
159-03, and 3C 295, e.g., see Fig. 1$c,d$) require unphysically large
values of $f>1$.  The latter five clusters exhibit strong AGN activity
in their centers and extended radio emission, which might indicate
that the gas receives extra heat energy from the AGN \cite{NK_zak03}.

The turbulent mixing model fits all ten clusters quite well, with a
surprisingly narrow range of $\alpha_{\rm mix}\sim0.01-0.03$
\cite{NK_kim03b}.  The five clusters that were incompatible with the
conduction model tend to need a larger value of $\alpha_{\rm mix}$ by
a factor of 2 than the other clusters (perhaps because the nuclear
activity and the associated jets in these clusters cause enhanced
turbulent transport).  The values of $\alpha_{\rm mix}$ found from the
model fitting correspond to a turbulent diffusion coefficient of
$\kappa_{\rm mix}\sim1-6$ kpc$^2$ Myr$^{-1}$ at $r\sim50-300$ kpc,
which is similar to the value one infers from typical parameters for
intracluster turbulence: turbulent velocities $v_{\rm turb}\sim
100-300$ km s$^{-1}$ and eddy sizes $l_B\sim5-20$ kpc
\cite{NK_ric01,NK_car02}.

\section{Thermal Stability}\label{NK_stab}

Since optically-thin gas at X-ray temperatures is known to be
thermally unstable, it is necessary to check the stability of the
equilibrium models discussed in \S3.  The absence of cold material in
the centers of clusters indicates that the thermal instability is
either absent or at least very weak.  Since diffusive processes in
general tend to stabilize thermal instability on small scales
\cite{NK_fie65}, it is interesting to ask whether thermal conduction
with $f\sim 0.2-0.4$ or turbulent mixing with $\alpha_{\rm
mix}\sim0.01-0.03$ can suppress the growth of large-scale unstable
modes in clusters.

We begin with a discussion of local linear modes, where we assume that
the perturbations have rapid spatial variations.  It is
straightforward to derive a dispersion relation for such modes.  Using
equation (\ref{NK_Flux}) for the total heat flux, we find
\begin{equation}\label{NK_sig}
\sigma  = \sigma_\infty - \kappa_{\rm mix} ( 1 + q) k_r^2,
\end{equation}
where $\sigma$ is the growth rate of the model, $\sigma_\infty\equiv
3(\gamma-1)j/(\gamma P)$ is the growth rate of isobaric perturbations
in the absence of diffusion \cite{NK_kim03a,NK_fie65}, $k_r$ is the
radial wavenumber of the mode, and the dimensionless parameter
$q\equiv(\gamma-1)\kappa_{\rm cond}T/ (\gamma\kappa_{\rm mix}P)$
measures the stabilizing effect of conduction relative to mixing.
Putting in numerical values, clusters with pure conduction should be
marginally stable to local perturbations \cite{NK_zak03}.  Since $q
\sim 0.1 (f/0.2)(0.02/\alpha_{\rm mix}) (r/20\,{\rm kpc})^{-1}
(n_e/0.05\,{\rm cm^{-3}})^{-1}$ is normally less than unity in the
region $r<20$ kpc where most of the cooling occurs, we expect
turbulent mixing to have a stronger stabilizing effect relative to
conduction.

We have confirmed these predictions by explicitly analyzing the global
stability of the equilibrium models.  By applying Lagrangian
perturbations and solving the perturbed equations as a boundary value
problem, we searched for all unstable/overstable modes and calculated
their growth times $t_{\rm grow}$.  In the presence of conduction, we
find that all global modes become stable except for the fundamental,
nodeless mode.  The lone unstable mode has a very long growth time,
e.g., A1795 with $f=0.2$ has $t_{\rm grow}\sim4.1$ Gyr, while Hydra A
with $f=3.5$ has $t_{\rm grow}\sim9.3$ Gyr \cite{NK_kim03a}.
Turbulent mixing suppresses the instability even more significantly;
A1795 with $\alpha_{\rm mix}=0.011$ has $t_{\rm grow}$ much longer
than the Hubble time, and Hydra A with $\alpha_{\rm mix}=0.021$ is
completely stable \cite{NK_kim03b}.  These results suggest that
thermal instability is not a serious issue for clusters that achieve
thermal balance through diffusive heat transport.

\section{Scaling Laws}

The theory of cosmic structure formation indicates that the mass $M$
of a halo should scale with the virial temperature $T$ as $M\propto
T^{3/2}$, and that the X-ray luminosity and the entropy should scale
as $L_X\propto T^2$ and $S\equiv Tn_e^{-2/3}\propto T$.  However,
cluster observations show different scaling laws: $M\propto
T^{1.7\sim1.9}$, $L_X\propto T^{2.5\sim3}$, $S\propto T^{0.6\sim0.7}$,
for rich clusters with $kT{\lower.5ex\hbox{$\; \buildrel > \over \sim
\;$}} 2$ keV \cite{NK_all98,NK_san03,NK_pon03}; and $L_X\propto
T^{4\sim5}$, $S\propto T^{-0.7\sim 0.2}$, for small clusters or galaxy
groups with $kT{\lower.5ex\hbox{$\; \buildrel < \over \sim \;$}} 1$
keV \cite{NK_pon03,NK_hel00}.  That is, not only are the observed
power-law indices different from the self-similar predictions, there
is also a clear break in cluster properties at a characteristic
temperature $kT\sim1-2$ keV.  The fact that smaller clusters or groups
have relatively constant entropy has been recognized as an ``entropy
floor.''  The prevailing explanations for the rather high entropy at
low temperatures include pre-heating of intracluster gas
\cite{NK_kai91,NK_evr91}, removal of cold low-entropy gas via galaxy
formation in clusters \cite{NK_bry00}, and supernova feedback
\cite{NK_voi02}.  Although some of these suggestions are fairly
successful in reproducing the entropy floor and the observed scalings,
none of them includes thermal conduction or turbulent mixing.  If
these processes are at all important in clusters, they should have a
large effect on the scaling laws.

It is straightforward to derive scaling relationships that the
equilibrium cluster models of \S3 should obey.  For rich clusters with
$kT{\lower.5ex\hbox{$\; \buildrel > \over \sim \;$}} 2$ keV, where
thermal bremsstrahlung ($j\propto n_e^2T^{1/2}$) dominates, heating by
conduction leads to $L_X\propto T^4$ and $S\propto T^{0.3}$, while
heating by turbulent mixing predicts $L_X\propto T^3$ and $S\propto
T^{0.6}$.  On the other hand, for small clusters or groups
($kT{\lower.5ex\hbox{$\; \buildrel < \over \sim \;$}} 1$ keV), where
cooling is dominated by line transitions ($j\propto
n_e^2T^{-0.7\sim-1}$), $L_X\propto T^4$ and $S\propto T^{-0.2\sim0}$
for the thermal conduction model, and $L_X\propto T^{4.2\sim4.5}$ and
$S\propto T^{-0.3\sim-0.1}$ for the turbulent mixing model
\cite{NK_kim03b}.  We see that the scaling relations predicted by the
mixing model are in remarkably good agreement with the observations.
The dramatic change of cluster properties at $kT\sim(1-2)$ keV arises
because of the change in the cooling mechanism above and below this
temperature.  Also, the entropy floor observed in groups is reproduced
naturally.

\section{Conclusion}

The thermal conduction and turbulent mixing models have certain
attractive properties which ultimately are due to the fact that both
models involve diffusive transport.  Diffusion not only allows heat to
move into the cluster center from the outside, it also irons out
perturbations and thereby helps to control thermal instability.  What
is interesting is that the amount of diffusion required to fit the
observations is comparable to that predicted by theoretical arguments.

Two caveats need to be mentioned.  First, the presence of cold fronts
in many clusters \cite{NK_mar00,NK_vik01} indicates that large
temperature and entropy jumps are able to survive in some regions of
the hot gas.  Diffusion is clearly suppressed across these surfaces.
It is possible that cold fronts are special regions where the magnetic
field is combed out parallel to the front, thereby suppressing
cross-field conduction temporarily \cite{NK_vik01,NK_zak03}.

Second, all we have shown is that a cluster with the observed density
and temperature profile would be in hydrostatic and thermal
equilibrium and would be fairly stable.  However, we have not
explained how the cluster reaches the observed state starting from
generic initial conditions.  Time-dependent simulations show that a
cluster with thermal conduction would either slowly evolve to an
isothermal state if its initial density is less than a critical
density, or develop a catastophic cooling flow otherwise
\cite{NK_bre88}.  Does the current observed state result from an
initial rapid mass dropout (which decreases the density) and
subsequent slow evolution with diffusive heating of an once overdense
cluster \cite{NK_kim03a}? Are other heating mechanisms, e.g., AGNs,
necessary to explain the present state of clusters?  Answers to these
questions are of fundamental importance to understanding clusters and
more generally galaxy formation.

\vspace{0.5cm}
\noindent
{\bf Acknowledgments.} The work reported here was supported in part by
NASA grant NAG5-10780 and NSF grant AST 0307433.



\end{document}